# The effect of a coupling field on the entanglement dynamics of a three-level atom

Ali Mortezapour[1, 2], Majid Abedi[2], Mohammad Mahmoudi[1] and M R H Khajehpour[2]

[1] Physics Department, Zanjan University, P.O. Box 45195-313, Zanjan, Iran
[2] Institute for Advanced Studies in Basic Sciences, P.O. Box 45195-159, Zanjan, Iran
Email: a.mortezapour7@gmail.com
Email: mahmoudi@iasbs.ac.ir

**Abstract**. The effect of a coupling laser field on the entanglement of a three-level quantum system and its spontaneous emission is investigated via reduced quantum entropy. We consider two schemes, the upper- and lower-level couplings. By calculating the degree of entanglement (DEM) for both systems, it is shown that the entanglement between atom and its spontaneous emission can be controlled by the coupling laser field. This field, however, affects the entanglement differently in the two schemes; it is only the lower-level coupling scheme that shows a nonzero steady state DEM which can be controlled by the intensity and detuning of the coupling laser field.

## 1. Introduction

The three-level "atoms" interacting with a single-mode field or a pair of fields with arbitrary detuning have been a topic of interest to physicists for more than three decades. This has been mainly due to the fact that they are the simplest systems which display many nonlinear phenomena when irradiated with monochromatic electromagnetic fields. The three configurations $\Lambda$, V and $\Xi$ of these systems behave differently under applied coherent fields as well as in the processes involving spontaneous emission. In one of these phenomena, the coherent trapping, the application of the fields pumps the system in a coherent superposition state which is non-absorbing. The theoretical studies of Arimondo and Orriols [1] showed that the phenomenon is due to the presence of interference processes. The coherent population trapping (CPT) and related phenomena of electromagnetically induced transparency (EIT), enhancement of index of refraction, lasing without inversion (LWI), adiabatic population transfer etc [2] are all basically consequences of quantum coherence and interference. The early works in the field were reviewed by Yoo and Eberly [3]; Ariomondo's excellent review covers both theoretical and experimental works on coherent population trapping [4]. For a unified treatment of CPT, EIT, and LWI in a simple three-level model, as well as an extensive list of references see Scully and Zubairy [5]. An important step in the dissipative dynamics of three-level atoms was taken by Agarwal who showed that an indirect coupling between the excited states, in a V configuration, may develop if the dipole moments of transitions to ground state are parallel [6]. This indirect coupling is a consequence of interaction with vacuum. Hegerfeld and Plenio [7] studied the same configuration when the separation of uncoupled states was small and a resonant coherent field was applied. The system decayed rapidly from excited states to ground state, and for a suitable choice of the coherent field a surprising effect was obtained which resembled EIT and non-absorption resonances studied by Orriole [8] and Cardimona et al [9]: the atom stopped rapidly to fluoresce though still being irradiated. In a related work ultra-sharp emission lines were obtained in resonance fluorescence of a V-type system with almost parallel moments [10]. The case with perpendicular transition dipole moments also behave similarly when the ground state is coupled by a single-mode laser field to one of the excite states and the excited states are themselves coupled by a DC field [11]. In all the above systems the non-absorption, the ultra-narrowing of spectral lines and other effects are consequences of quantum interference between two transition pathways. Zhu and Scully's work on Autler-Townes spontaneous spectrum in a three-level atom demonstrated the same mechanism at work [12]. Of the two cases considered by them (upper-level coupling and lower-level coupling) only the one with the two upper levels coupled by a coherent field displayed the existence of dark state and the narrowing of the spectral peaks. Spontaneous emission in three-level atoms driven by coherent fields can also be used to study the strong correlations between the atom and its spontaneous fields. This so-called atom-field entanglement is a specifically quantum phenomenon in bipartite and multipartite systems. A system in an entangled state would have its subsystems lost their individuality and quoting Schrodinger "the whole [system] is in a definite state whereas the parts are not". The concept of entanglement and the name go back to the famous EPR article and Schrodinger's rejoinder [13] in 1935 on the foundations of quantum measurement. It is nowadays a key concept in quantum computation and quantum information theory [14]. The realization of entangled states in atom-field systems, NMR, trapped ions, quantum dots etc, provides us with new practical applications in diverse fields such as improved optical frequency standards [15] and quantum computer protocols.

In another line of development the three-level atoms interacting with bimodal detuned cavities have been used to present schemes for the realization of universal two-qubit logic gates such as control-NOT, QPG, and SWAP, on which the performance of any quantum computer depends [16].

Entanglement between a Λ-type three-level atom and its two spontaneous emission fields is studied using the quantum entropy. A steady state entanglement between the atom and its spontaneous fields is obtained while the correlation of two spontaneous fields is shown to be classical [17]. The effect of a classical coupling laser field on the entanglement of a Ξ-type three-level atom and its two successive spontaneous emissions is investigated using the same approach and again a steady state entanglement of the system is reported; the coupling laser field here couples the ground state and the higher excited state [18]. The effect of quantum interference of spontaneous emission on the entanglement of a driven V-type three-level atom and its spontaneous emission has recently been studied and the disentanglement behavior of atom-photon system via quantum interference is reported [19].

In this paper we study the effect of a coupling laser field on the entanglement of a three level atom and its spontaneous emission in two schemes of upper-level and lower-level coupling using quantum entropy. Both schemes are treated within the electric dipole approximation and RWA is used all throughout. The model was originally used by Zhu and Scully [12] to study the interference effects. There exist some plausible realizations for both upper and lower level schemes which will be discussed briefly in the concluding section. We have worked out the time evolution of atomic reduced entropy and the entanglement of the atom and its spontaneous emission in both schemes. It is demonstrated that the effect of the coupling laser field on entanglement in the two schemes are different. Such a differentiation is not available in the system studied in reference [18] where two successive spontaneous emissions are involved. An important result obtained is that the steady state entanglement of the atom and its spontaneous emission can be controlled by the intensity and detuning of the coupling laser field, only for the case of lower-level coupling. In the upper-level coupling, the steady state entanglement of the atom and its spontaneous emission vanishes for all intensities. The coupling laser field is treated both as a classical and a quantized field though the calculations are presented mostly for the quantized one.

## 2. Model and Solution

The model under consideration is a three-level quantum atom interacting with a coherent coupling laser field. The system undergoes a single spontaneous transition from one of the higher levels to the ground state, so a coupling to the vacuum modes is assumed. Depending on which two levels are coupled by the coupling laser field two schemes are considered, namely upper-level and lower-level coupling, which are shown in figure 1(a and b). Atomic states of the system are denoted by $|a\rangle$, $|b\rangle$ and $|c\rangle$.

### *2.1 Upper-level Coupling*

Here the upper level $|c\rangle$ of the atom is coupled resonantly to level $|a\rangle$ by a coherent field and the transition from $|a\rangle$ to ground state $|b\rangle$ is coupled to vacuum modes (see figure 1(a)). If the coupling laser field is treated as a classical field, the Hamiltonian of the system, in the interaction picture can be written as

$$V = \Omega|c\rangle\langle a|e^{i(\omega_{ca}-\omega_l)t} + \sum_k g_k|a\rangle\langle b|\hat{b}_k e^{i(\omega_{ab}-\omega_k)t} + \text{H.c.}, \qquad (1)$$

where $\omega_{ca}$, $\omega_{ab}$ are the frequencies of $|c\rangle \leftrightarrow |a\rangle$ and $|a\rangle \to |b\rangle$ transitions, respectively; $b_k^\dagger$ ($b_k$) is creation (annihilation) operator for the $k$ th vacuum mode with frequency $\omega_k$; $g_k$ denotes the coupling constant between the $k$ th vacuum mode and the atomic transition $|a\rangle \to |b\rangle$. The parameters $\omega_l$ and $\Omega$ are the frequency and the Rabi frequency of the coupling laser field. We set $\hbar = 1$ and take $g_k$ as real.

We assume the atom to be initially in a pure superposition of $|a\rangle$ and $|c\rangle$,

$$|\Psi_A(0)\rangle = C(0)|c\rangle + A(0)|a\rangle, \qquad (2)$$

where $|C(0)|^2 + |A(0)|^2 = 1$, and the field in state $|0\rangle$. At any later time $t$, the quantum state of the system is described by

$$|\Psi_{AF}(t)\rangle = C(t)|c\rangle|0\rangle + A(t)|a\rangle|0\rangle + \sum_k B_k(t)|b\rangle|1_k\rangle, \qquad (3)$$

where $|1_k\rangle$ denotes the state with one photon in $k$ th vacuum mode. Substituting Eqn (3) into the Schrödinger equation we obtain the equations of motion for the probability amplitudes $C(t)$, $A(t)$ and $B_k(t)$,

$$\dot{C}(t) = -i\Omega A(t) e^{i\Delta t}, \qquad (4a)$$

$$\dot{A}(t) = -i\Omega^* C(t) e^{-i\Delta t} - i\sum_k g_k B_k(t) e^{i\delta_k t}, \qquad (4b)$$

$$\dot{B}_k(t) = -ig_k A(t) e^{-i\delta_k t}, \qquad (4c)$$

where $\delta_k = \omega_k - \omega_{ab}$ and $\Delta = \omega_{ca} - \omega_l$.

Solving the above equations, while taking the Weisskopf–Wigner theory into account, the amplitudes are obtained as

$$C(t) = C^{(1)} e^{x_1 t} + C^{(2)} e^{x_2 t}, \qquad (5a)$$

$$A(t) = A^{(1)} e^{y_1 t} + A^{(2)} e^{y_2 t}, \qquad (5b)$$

$$B_k(t) = -ig_k [A^{(1)} \frac{e^{(y_1+i\delta_k)t}-1}{y_1+i\delta_k} + A^{(2)} \frac{e^{(y_2+i\delta_k)t}-1}{y_2+i\delta_k}], \qquad (5c)$$

where, defining $\alpha = \frac{\gamma}{2} + i\Delta$ and $\beta = [\alpha^{*2} - 4|\Omega|^2]^{1/2}$,

$$x_{1,2} = -(\alpha^* \pm \beta)/2,$$

$$y_{1,2} = -(\alpha \pm \beta)/2, \tag{6}$$

and,

$$A^{(1)} = -[A(0)(y_1 + i\Delta) - iC(0)\Omega]/\beta, \qquad A^{(2)} = A(0) - A^{(1)},$$

$$C^{(1)} = -[C(0)(x_1 + \alpha^*) - iA(0)\Omega^*]/\beta, \qquad C^{(2)} = C(0) - C^{(1)}. \tag{7}$$

Here $\gamma$ is the spontaneous emission rate from level $|a\rangle$ to level $|b\rangle$.

Now we treat the coupling laser field as a quantized field. The initial state of the atom can be written as Eqn (2). For the coupling field we have initially,

$$|\Psi_l(0)\rangle = \sum_n w_n |n\rangle \tag{8}$$

Where $|n\rangle$ are the number states and $w_n = e^{-m/2} m^{n/2} e^{in\theta} / \sqrt{n!}$ defines the coherent distribution of the number states, $m$ is the mean value of the photon number in the field and $\theta$ is the phase, set for simplicity equal to zero. The Hamiltonian of the system, in interaction picture can be written as

$$V = g|c\rangle\langle a|\hat{a} e^{i(\omega_{ca} - \omega_l)t} + \sum_k g_k |a\rangle\langle b|\hat{b}_k e^{i(\omega_{ab} - \omega_k)t} + \text{H.c.}, \tag{9}$$

where $\hat{a}(\hat{a}^\dagger)$ is the annihilation (creation) operator of photons with frequency $\omega_l$ and g is the coupling of the atom and the laser field.

The time dependent quantum state of the of the atom-field system is described by

$$|\Psi_{AF}(t)\rangle = \sum_n \{C_n(t)|c, n-1\rangle|0\rangle + A_n(t)|a, n\rangle|0\rangle + \sum_k B_{n,k}(t)|b, n\rangle|1_k\rangle\}. \tag{10}$$

Where $|a, n\rangle$ denotes the atomic state $|a\rangle$ with $n$ photons in the coupling laser field and so on with other states. Substituting Eqn (10) into the Schrodinger equation we obtain the equations of motion for probability amplitudes,

$$\dot{C}_n(t) = -ig\sqrt{n} A_n(t) e^{i\Delta t}, \tag{11a}$$

$$\dot{A}_n(t) = -ig^*\sqrt{n} C_n(t) e^{-i\Delta t} - i\sum_k g_k B_{n,k}(t) e^{i\delta_k t}, \tag{11b}$$

$$\dot{B}_{n,k}(t) = -ig_k A_n(t) e^{-i\delta_k t}. \tag{11c}$$

The probability amplitudes $A_n(t)$, $C_n(t)$ and $B_{n,k}(t)$ are obtained as before by integrating Eqns (11) using Weisskopf–Wigner theory,

$$C_n(t) = w_{n-1}[C_{n-1}^{(1)} e^{x_1(n-1)t} + C_{n-1}^{(2)} e^{x_2(n-1)t}], \tag{12a}$$

$$A_n(t) = w_n[A_n^{(1)} e^{y_1(n)t} + A_n^{(2)} e^{y_2(n)t}], \tag{12b}$$

$$B_{n,k}(t) = -ig_k[A_n^{(1)} \frac{e^{[y_1(n)+i\delta_k]t} - 1}{y_1(n) + i\delta_k} + A_n^{(2)} \frac{e^{[y_2(n)+i\delta_k]t} - 1}{y_2(n) + i\delta_k}], \tag{12c}$$

where, defining $\beta(n) = [\alpha^{*2} - 4|g|^2(n)]^{1/2}$,

$$x_{1,2}(n) = -[\alpha^* \pm \beta(n)]/2,$$

$$y_{1,2}(n) = -[\alpha \pm \beta(n)]/2,$$

$$A_n^{(1)} = -\{A_n(0)[y_1(n) + i\Delta] - iC_n(0)g\sqrt{n}\}/\beta(n), \qquad A_n^{(2)} = A_n(0) - A_n^{(1)},$$

$$C_n^{(1)} = -\{C_n(0)[x_1(n) + \alpha^*] - iA_n(0)g^*\sqrt{n}\}/\beta(n), \qquad C_n^{(2)} = C_n(0) - C_n^{(1)}. \tag{13}$$

Comparing $\beta$ and $\beta(n)$ we see that for each $n$, $|\Omega|^2 \to |g|^2 n$. But the averaging process over all $n$ enters in a complicated way and one should expect that any final result will be different than the simple correspondence $|\Omega|^2 \to |g|^2 m$, except for large values of photon number.

We may rewrite the state vector Eqn (10) of the quantum system as:

$$|\Psi(t)\rangle = |P\rangle|c\rangle + |Q\rangle|a\rangle + |R\rangle|b\rangle \tag{14}$$

Where

$$|P\rangle = \sum_n C_n(t)|n-1\rangle|0\rangle,$$

$$|Q\rangle = \sum_n A_n(t)|n\rangle|0\rangle,$$

$$|R\rangle = \sum_n \sum_k B_{n,k}(t)|n\rangle|1_k\rangle. \tag{15}$$

*2.2 Lower-level coupling*

In lower-level coupling scheme, figure 1(b), level $|c\rangle$ is coupled to level $|b\rangle$ by a classical coupling laser field of frequency $\omega_l$. The transition from $|a\rangle$ to $|c\rangle$ is not allowed, and the system decays directly to ground state from state $|a\rangle$. The interaction Hamiltonian for the lower-level coupling is

$$V = \Omega|c\rangle\langle b|e^{i(\omega_{cb} - \omega_l)t} + \sum_k g_k|a\rangle\langle b|\hat{b}_k e^{i(\omega_{ab} - \omega_k)t} + \text{H.c.} \tag{16}$$

Regarding the coupling of $|c\rangle$ and $|b\rangle$ it is convenient to use the dressed states.

$$|\chi_+\rangle = \varepsilon|c\rangle + \eta|b\rangle, \qquad |\chi_-\rangle = -\eta|c\rangle + \varepsilon|b\rangle, \tag{17}$$

where

$\varepsilon = \lambda_1 / \sqrt{\lambda_1^2 + |\Omega|^2}$, $\eta = \Omega / \sqrt{\lambda_1^2 + |\Omega|^2}$ and $\lambda_{1,2} = (\Delta' \pm \sqrt{\Delta'^2 + 4|\Omega|^2})/2$.

Here $\lambda_1$ and $\lambda_2$ are the corresponding eigenvalues of $|\chi_+\rangle$ and $|\chi_-\rangle$, and $\Delta' = \omega_{cb} - \omega_l$ the detuning of the coupling laser field. We can write the Hamiltonian in the interaction picture as

$$V_I = \sum_k g_k [\eta |a\rangle\langle \chi_+ | \hat{b}_k e^{-i(\delta_k + \lambda_1)t} + \varepsilon |a\rangle\langle \chi_- | \hat{b}_k e^{-i(\delta_k + \lambda_2)t}] + H.c., \tag{18}$$

where $\delta_k = \omega_k - \omega_{ab}$.

We assume the system to be initially in a product state with the atom in $|a\rangle$ and the emission field in its vacuum state,

$$|\Psi_{AF}(0)\rangle = |a\rangle \otimes |0\rangle \tag{19}$$

At any time t, the quantum state of lower-level coupling system is described by

$$|\Psi_{AF}(t)\rangle = A(t)|a\rangle|0\rangle + \sum_k [X_k^+(t)|\chi_+\rangle|1_k\rangle + X_k^-(t)|\chi_-\rangle|1_k\rangle]. \tag{20}$$

The equations of motion of probability amplitudes $A(t)$, $X_k^+(t)$ and $X_k^-(t)$ will then be given by,

$$\dot{A}(t) = -i \sum_k g_k [X_k^+(t)\eta e^{-i(\delta_k + \lambda_1)t} + X_k^-(t)\varepsilon e^{-i(\delta_k + \lambda_2)t}], \tag{21a}$$

$$\dot{X}_k^+(t) = -i g_k \eta A(t) e^{i(\delta_k + \lambda_1)t}, \tag{21b}$$

$$\dot{X}_k^-(t) = -i g_k \varepsilon A(t) e^{i(\delta_k + \lambda_2)t}. \tag{21c}$$

We can obtain the probability amplitudes for lower-level coupling, in Weisskopf–Wigner approximation,

$$A(t) = A(0) e^{-\frac{\gamma}{2}t}, \quad A(0) = 1,$$

$$X_k^+(t) = i g_k \eta \frac{[1 - e^{i(\delta_k + \lambda_1)t - \frac{\gamma t}{2}}]}{i(\delta_k + \lambda_1) - \frac{\gamma}{2}},$$

$$X_k^-(t) = i g_k \varepsilon \frac{[1 - e^{i(\delta_k + \lambda_2)t - \frac{\gamma t}{2}}]}{i(\delta_k + \lambda_2) - \frac{\gamma}{2}}, \tag{22}$$

where $\gamma$ denotes the spontaneous emission rate from level $|a\rangle$ to level $|b\rangle$.

We may again rewrite the state vector of the system as:

$$|\Psi(t)\rangle = |P\rangle|a\rangle + |Q\rangle|\chi_+\rangle + |R\rangle|\chi_-\rangle, \tag{23}$$

where

$$|P\rangle = A(t)|0\rangle, \quad |Q\rangle = \sum_k X_k^+(t)|1_k\rangle \text{ and } |R\rangle = \sum_k X_k^-(t)|1_k\rangle. \tag{24}$$

Now we assume that the transition $|c\rangle \leftrightarrow |b\rangle$ is excited by a quantized coupling laser field. The interaction Hamiltonian can be written as

$$V = g|c\rangle\langle b|\hat{a}e^{i(\omega_{cb}-\omega_l)t} + \sum_k g_k|a\rangle\langle b|\hat{b}_k e^{i(\omega_{ab}-\omega_k)t} + \text{H.c.} \tag{25}$$

We use the following dressed states whose corresponding eigenvalues are $\lambda_{1,2}(n)$:

$$|\chi_+, n\rangle = \varepsilon(n)|c, n-1\rangle + \eta(n)|b, n\rangle, \quad |\chi_-, n\rangle = -\eta(n)|c, n-1\rangle + \varepsilon(n)|b, n\rangle, \tag{26a}$$

where

$$\varepsilon(n) = \lambda_1(n)/\sqrt{\lambda_1^2(n) + n|g|^2}, \qquad \eta(n) = g\sqrt{n}/\sqrt{\lambda_1^2(n) + n|g|^2},$$

and $\lambda_{1,2} = (\Delta' \pm \sqrt{\Delta'^2 + 4n|g|^2})/2$. \hfill (26b)

Then the Hamiltonian in the interaction picture can be written as

$$V_I = \sum_k g_k[\eta(n)|a,n\rangle\langle\chi_+,n|\hat{b}_k e^{-i(\delta_k+\lambda_1(n))t} + \varepsilon(n)|a,n\rangle\langle\chi_-,n|\hat{b}_k e^{-i(\delta_k+\lambda_2(n))t}] + H.c. \tag{27}$$

If we assume that the initial state of the system is given by a product state of the atom (in $|a\rangle$), the emission field (in vacuum $|0\rangle$) and the coupling laser field (in a coherent state given by Eqn (8)), the quantum state of the lower-level coupling system at time t is described by:

$$|\Psi_{AF}(t)\rangle = \sum_n \{A_n(t)|a,n\rangle|0\rangle + \sum_k [X_{n,k}^+(t)|\chi_+,n\rangle|1_k\rangle + X_{n,k}^-(t)|\chi_-,n\rangle|1_k\rangle]\}. \tag{28}$$

The probability amplitudes $A_n(t)$, $X_{n,k}^+(t)$ and $X_{n,k}^-(t)$ can be calculated, using the aforementioned methods, as

$$A_n(t) = A(0)w_n e^{-\frac{\gamma}{2}t}, \quad A(0) = 1,$$

$$X_{n,k}^+(t) = ig_k\eta(n)w_n \frac{[1-e^{i(\delta_k+\lambda_1(n))t - \frac{\gamma t}{2}}]}{i(\delta_k+\lambda_1(n)) - \frac{\gamma}{2}},$$

$$X_{n,k}^-(t) = ig_k\varepsilon(n)w_n \frac{[1-e^{i(\delta_k+\lambda_2(n))t - \frac{\gamma t}{2}}]}{i(\delta_k+\lambda_2(n)) - \frac{\gamma}{2}}. \tag{29}$$

## 3. Entropy and Entanglement

The state of a bipartite system composed of two distinguishable, sometimes coupled subsystems $A$ and $F$ (say atom and field) is described by a density operator $\rho_{AF}$ in a tensor product space $\mathcal{H}_A \otimes \mathcal{H}_F$. $\mathcal{H}_{A(F)}$ is the Hilbert space of the subsystem $A(F)$. The expectation value of any general variable in a composite system is determined by $\rho_{AF}$, but the expectation values of observables of the subsystem $A(F)$ are obtained using the reduced density operator,

$$\rho_{A(F)} = \text{Tr}_{F(A)} \rho_{AF} , \tag{30}$$

where the partial trace is taken over $\mathcal{H}_{F(A)}$.

The pure state $|\Psi\rangle_{AF}$ of the composite (atom-field) system in $\mathcal{H}_A \otimes \mathcal{H}_F$ is said to be separable if it can be written as the tensor product of $|\Psi\rangle_A$ and $|\Psi\rangle_F$ in $\mathcal{H}_A$ and $\mathcal{H}_F$ respectively,

$$|\Psi\rangle_{AF} = |\Psi\rangle_A \otimes |\Psi\rangle_F , \tag{31}$$

and hence the corresponding density operator as,

$$\rho_{AF} = \rho_A \otimes \rho_F \tag{32}$$

Otherwise the pure state $|\Psi\rangle_{AF}$ is called an entangled state [20]. A pure state of a bipartite system is entangled (or $A$ and $F$ are entangled) if and only if the reduced density operators for the subsystems describe mixed states. If reduced density operators describe pure states, i.e. $\rho_{A(F)}^2 = \rho_{A(F)}$, then the state of the system is separable. The entanglement of the mixed states of composite systems can also be defined on the basis of separability.

Defining useful measures of entanglement for general composite systems is a complicated problem and a hot topic of research at present time [21]. For the bipartite systems such as atom-field systems, there is a unique measure of entanglement under very general conditions (Schmidt decomposition, local invariance, continuity and additivity). This unique measure is the quantum or von Neumann entropy of reduced density operators. The von Neumann entropy S of a system in quantum state $\rho$ is defined as,

$$S = -\text{Tr} \, \rho \ln \rho . \tag{33}$$

It is the embodiment of total statistical properties of the quantum system as well as a powerful tool for investigation of its time evolution and dynamical properties. It obviously vanishes for any pure state and it is non-zero, $S \neq 0$, for mixed states. In a bipartite quantum system the system and subsystem entropies, at any time $t$ satisfy a remarkable inequality due to Araki and Lieb [22]:

$$|S_A(t) - S_F(t)| \leq S_{AF}(t) \leq S_A(t) + S_F(t), \tag{34}$$

where $S_{AF}(t) = -\text{Tr}\, \rho_{AF} \ln \rho_{AF}$ is the total entropy of the composite system and $S_{A(F)}(t) = -\text{Tr}_{A(F)}\, \rho_{A(F)} \ln \rho_{A(F)}$ are partial entropies corresponding to reduced density operators. Based on Eqn (34), for a closed atom-field system initially in a disentangled pure state, partial entropies of the field and the atom will be equal at all times after the interaction of two subsystems is switched on. Then our information about any of the subsystems is an indication of the entanglement of the whole system. A decreasing partial entropy means that each subsystem evolves towards a pure quantum state, whereas in an initially pure system an increasing partial entropy drives the two components to lose their individuality and become entangled [23]. So the degree of entanglement (DEM) for atom-field system would be,

$$DEM(t) = S_A = S_F = -\sum_{j=1}^{3} \lambda_j \ln \lambda_j, \tag{35}$$

where $\lambda_j$s are the eigenvalues of reduced density matrix of atom.

Now we go back to the system we studied in Section 2. The state vector $|\Psi_{AF}(t)\rangle$ of our atom-field system at time $t$ was given for different schemes by Eqns (10), (14), (23) and (28). The density operator of the system in all cases was of the form,

$$\rho_{AF} = |\Psi_{AF}(t)\rangle\langle\Psi_{AF}(t)| = \begin{pmatrix} |P\rangle\langle P| & |P\rangle\langle Q| & |P\rangle\langle R| \\ |Q\rangle\langle P| & |Q\rangle\langle Q| & |Q\rangle\langle R| \\ |R\rangle\langle P| & |R\rangle\langle Q| & |R\rangle\langle R| \end{pmatrix}. \tag{36}$$

Taking partial trace of $\rho_{AF}$ over all field variables, the reduced density matrix of the atom is obtained as

$$\rho_A = Tr_F(\rho_{AF}) = \begin{pmatrix} \langle P|P\rangle & \langle P|Q\rangle & \langle P|R\rangle \\ \langle Q|P\rangle & \langle Q|Q\rangle & \langle Q|R\rangle \\ \langle R|P\rangle & \langle R|Q\rangle & \langle R|R\rangle \end{pmatrix}. \tag{37}$$

For the upper-level scheme, with quantized coupling laser field,

$$\langle P|P\rangle = \sum_n |C_n(t)|^2$$

$$\langle Q|Q\rangle = \sum_n |A_n(t)|^2$$

$$\langle R|R\rangle = \sum_n \sum_k |B_{n,k}(t)|^2 = 1 - \langle Q|Q\rangle - \langle P|P\rangle \tag{38}$$

$$\langle P|Q\rangle = \sum_n C^*_{n+1}(t)A_n(t) = \langle Q|P\rangle^*$$

$$\langle P|R\rangle = \langle R|P\rangle^* = 0$$

$$\langle Q|R\rangle = \langle R|Q\rangle^* = 0$$

and for the lower-level scheme,

$$\langle P|P\rangle = \sum_n |A_n(t)|^2$$

$$\langle Q|Q\rangle = \sum_n \sum_k |\chi^+_{n,k}(t)|^2 = \sum_n |w_n|^2 |\eta(n)|^2 [1 - e^{-\gamma t}]$$

$$\langle R|R\rangle = \sum_n \sum_k |\chi^-_{n,k}(t)|^2 = \sum_n |w_n|^2 |\varepsilon(n)|^2 [1 - e^{-\gamma t}] \tag{39}$$

$$\langle P|Q\rangle = 0$$

$$\langle R|Q\rangle = \langle Q|R\rangle = \sum_n \frac{|w_n|^2 \gamma \varepsilon^*(n)\eta(n)[1 - e^{i(\lambda_1(n)-\lambda_2(n))t - \gamma t}]}{\gamma - i(\lambda_1(n)-\lambda_2(n))}.$$

The matrix elements of the reduced density operator for the cases of classical coupling laser field are calculated using Eqns (3) and (20).

## 4. Results and Conclusion

In this section, we shall first suggest two plausible realizations of the model, one for the upper level and the other for the lower-level scheme. Then, the results of our calculations and the conclusions are discussed.

Consider the levels of a calcium atom (figure 1(c)). The ground state is a $4s^2\ ^1S_0$ level. For the upper level scheme the excited state $4s6p\ ^1P_1$ can be coupled to the metastable state $4s3d\ ^1D_2$ by a field of amplitude $E_{ac}$ and wavelength 504 nm. The corresponding Rabi frequency would be $\Omega = (\vec{d}_{ac}\cdot\vec{E}_{ac})/\hbar$, where $d_{ac}$ is the transition dipole moment. The rate $\gamma_{ac}$ of spontaneous decay from the excited state to metastable state is negligible compared to the decay rate $\gamma$ from the excited state to the ground state. For the standard atomic parameters of the system [24-26] we have $\gamma_{ac} \approx 79$ Hz, $\gamma \approx 48$ MHz, $\hbar\omega_{ac} = 4\times 10^{-19}$ J, and $d_{ac} = 9.5 \times 10^{-30}$ Cm. The amplitude of the coupling laser field is obtained as $E_{ac} = 530$ V/m (field intensity $I_{ac} = 3.7 \times 10^{-2}$ W/cm$^2$) for $\Omega = \gamma$. For the lower-level coupling case, the ground state $4s^2\ ^1S_0$ can be coupled to the state $4s4p\ ^3P_1$ by a diode laser of wavelength 657 nm. The corresponding decay rate is $\gamma_{cb} \approx 2300$ Hz, while the decay of the excited level $4s4p\ ^1P_1$ to the ground state occurs at the much faster rate of $\gamma \approx 216$ MHz.

It is to be noted that the above upper level scheme is slightly different than the one in figure 1(a), namely, the coupled levels are interchanged. This interchange, however, has no effect on the reduced atomic entropy of the system.

We have assumed that the three-level system and its spontaneous emission are initially in a disentangled pure state. So the total entropy of the system remains equal to zero while it evolves. We have investigated the dynamic behavior of reduced atomic entropy for upper-level and lower-level coupling schemes. In the figures all frequencies and detunings are scaled to decay rate $\gamma$. The dynamic behavior of the atomic reduced entropy for different intensities of the coupling laser field is shown in figure 2 for upper-level coupling scheme. Figure 2(a) corresponds to the case when the coupling laser field is a classical one, and figure 2(b) is when it is a quantized field. It shows that the results for classical and quantized fields are in good agreement for large photon numbers; but for the low photon numbers they are somewhat different, as expected. The atomic reduced entropy of the system tends to zero in long time evolution, for all intensities of coupling laser field, i.e. the atom and its spontaneous emission eventually disentangle as $t \rightarrow \infty$. The physics of phenomenon can be described using the populations of levels shown in figure 3. Here the time evolution of populations of levels for different intensities of coupling laser field is shown. A close look at figure 3 shows that for different intensities, there is a transfer of population from levels $|c\rangle$ and $|a\rangle$ to ground state $|b\rangle$, though initially there is a slight increase in population of $|c\rangle$ at expense of $|a\rangle$. Level $|a\rangle$ depletes much faster than $|c\rangle$.

An interesting result is obtained in lower-level coupling scheme. In figure 4 we display the time evolution of atomic reduced entropy for different intensities of the coupling laser field. Figure 4(a) corresponds to a classical field coupling two states while figure 4(b) is for a quantized coupling field. We have assumed the atom to be initially in the excited state; the other parameters are the same as in figure 2. According to figure 4, a finite non-zero steady state value is obtained for the atomic entropy. Atom-field system in this scheme remains entangled and the degree of entanglement depends on intensity of the coupling laser field. The field is a handle with which one can control the entanglement.

Figure 5 shows the dynamical behavior of the population of dressed states in the lower-level coupling system for different intensities. It is plotted using the same parameters of figure 4. The upper level $|a\rangle$ decays to the ground state and then becomes empty. The coupling laser field re-distributes the probability amplitude between two lower levels. By changing the intensity of coupling laser field, the population amplitude of levels is changed and consequently the atomic entropy is also varied. As we increase the intensity of the coupling laser field the population of lower level will be transferred to the intermediate one. At high intensities the population of two lower dressed states (or two lower atomic levels) becomes equal. In this case the three-level system reduces to a simple two-level atomic system and the maximum value of atomic entropy, $Ln2$, is obtained for equal probability amplitude of energy levels.

Finally, the effect of detuning of the coupling laser field on the atomic reduced entropy is investigated. The steady state atomic entropy versus detuning for different values of coupling laser field is shown in figure 6. The maximum entropy is obtained in exact resonance.

In conclusion, we investigated the effect of coupling laser field on the dynamical behavior of reduced atomic reduced entropy in a three-level quantum system for two the upper- and lower-level couplings schemes. It is shown that the coupling laser field has different effects in two schemes. While the transient entanglement of atom and its spontaneous emission, in both schemes, can be changed by the coupling laser field, a controllable non-zero steady state atomic reduced entropy is obtained only in the case of lower-level coupling.

**Acknowledgment**

The authors gratefully acknowledge support by the Institute for Advanced Studies in Basic Science (IASBS) Research Council under grant No.G2010IASBS139.

**Figures caption**

Figure. 1. Proposed energy level schemes for (a) upper-level and (b) lower-level coupling scheme. The double arrow shows the coupling laser field. The wavy arrow denotes the spontaneous emission. The detuning of coupling laser field with respect to transition frequency is shown by $\Delta$ and $\Delta'$. Suggested realizations of the model for upper-level (Solid) and lower-level (Dotted) couplings are shown in (c).

Figure. 2 Dynamical behavior of atomic reduced entropy for different intensities of coupling laser field in upper-level coupling scheme. The coupling laser field is considered as a classical field with initial populations $C(0) = A(0) = 1/\sqrt{2}$, $B_k(0) = 0$ in (a), while it is considered as a quantized field with initial populations $C_n(0) = A_n(0) = 1/\sqrt{2}$, $B_{n,k}(0) = 0$ in (b). Parameters $\Delta, \Omega$ and $g$ are scaled to $\gamma$. $\Delta = 0.1$. In (a) $\Omega = 0.1$(Solid), 0.2 (Dotted), 1.0(Dashed). In (b) $g = 0.1$, $m = 100$(Solid); $g = 0.1$, $m = 4$ (Dotted); $g = 0.1$, $m = 100$(Dashed).

Figure. 3. Population of levels in upper-level coupling scheme, for different intensities of coupling laser field. The parameters are same as in Figure. 2 (a).

Figure. 4. Time evolution of atomic reduced entropy in lower-level coupling configuration, for different intensities of the coupling laser field. (a) corresponds to the case of classical laser field while (b) corresponds to a quantized coupling field. We assume the atom is initially in the excited state and other parameters are scaled to $\gamma = 1$, $\Delta' = 0.1$. In (a) $\Omega = 0.1$(Solid), 0.2 (Dotted), 1.0(Dashed), 0.5(Dash-dotted). In (b) $g = 0.1$, $m = 100$ (Solid); $g = 0.1$, $m = 4$ (Dotted); $g = 0.1$, $m = 100$ (Dashed); $g = 0.5$, $m = 100$(Dash-dotted).

Figure. 5. Population of levels in lower-level coupling scheme, for different intensities of coupling laser field. The parameters are the same as in Figure. 4 (a).

Figure. 6. The steady state atomic reduced entropy versus detuning for different values of coupling laser field intensity. Initial amplitudes are $A(0) = 1$, $X^+(0) = X^-(0) = 0$ and $\Omega = 0.1$(Solid), 1.0(Dotted), 5.0(Dashed).

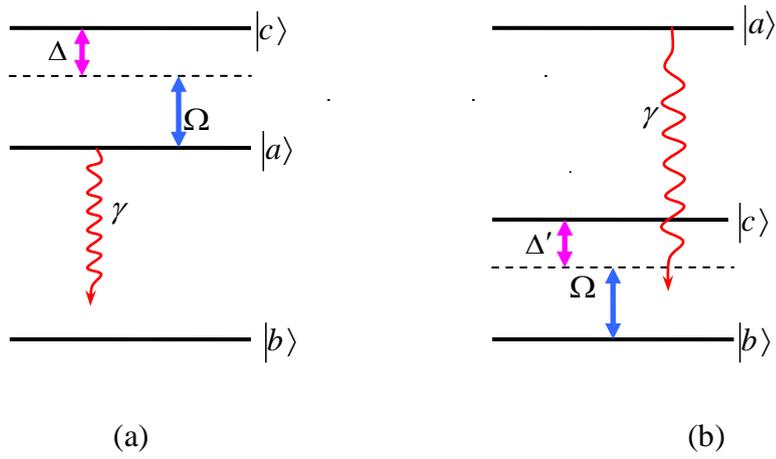

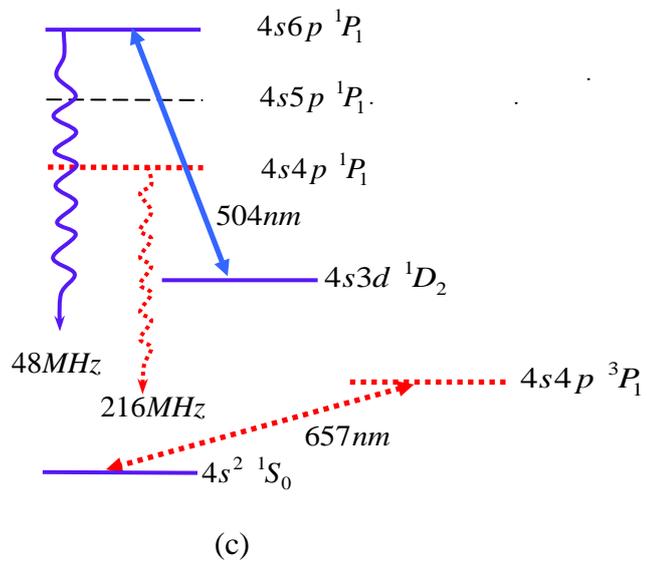

Figure 1

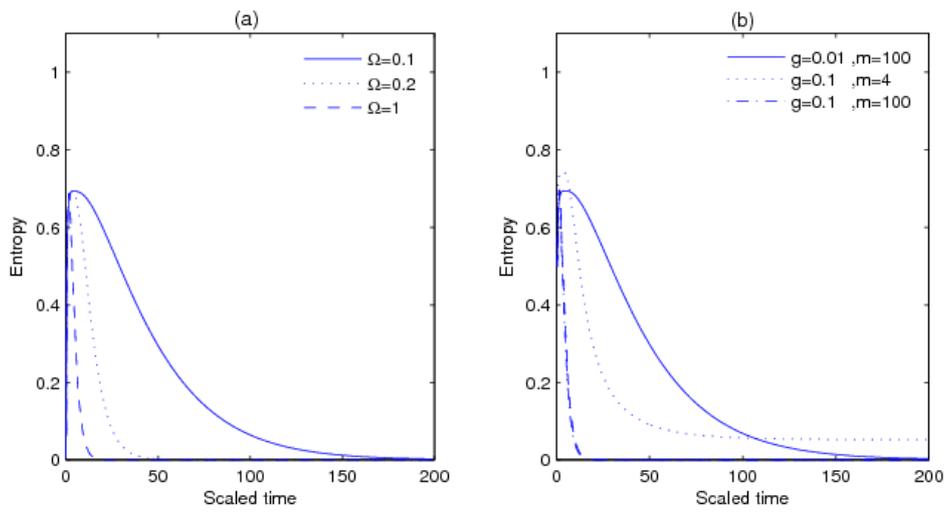

Figure 2

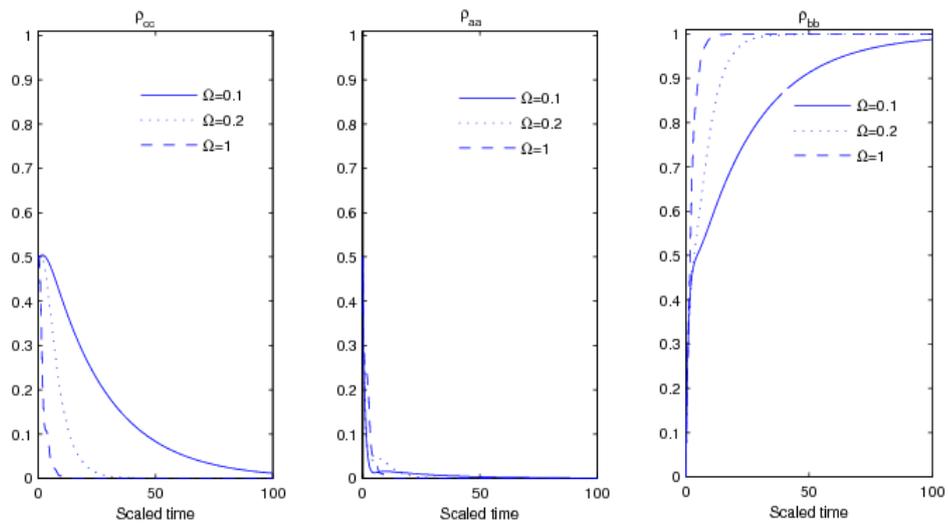

Figure 3

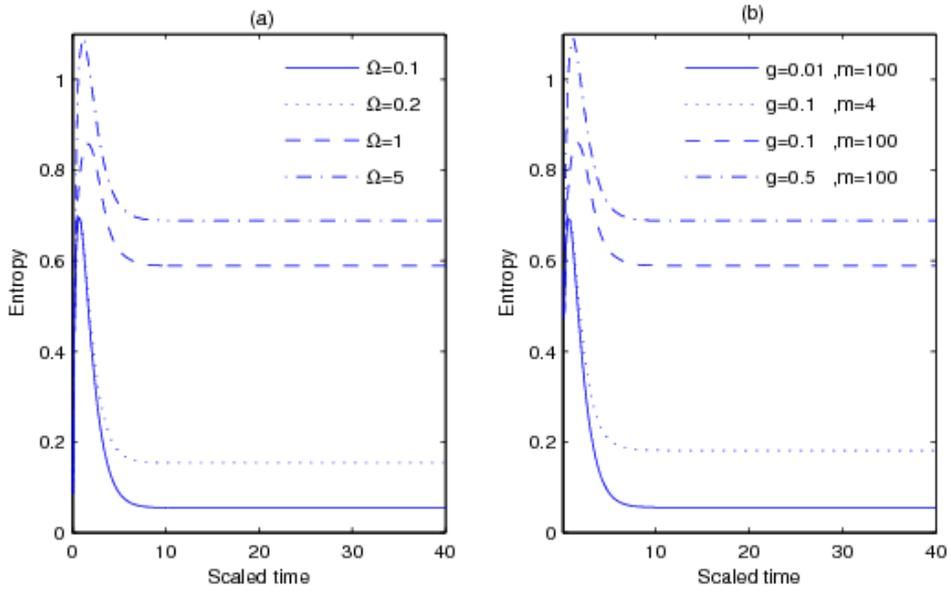

Figure 4

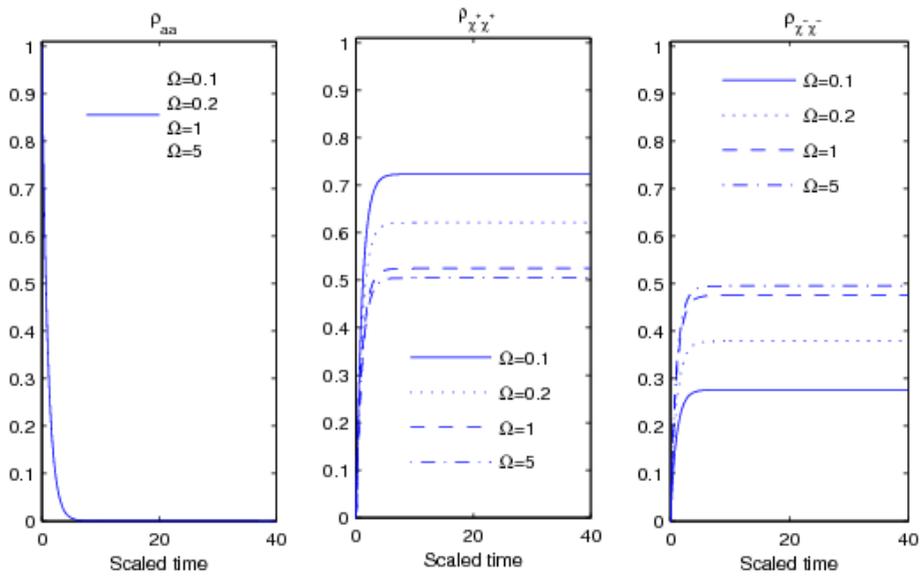

Figure 5

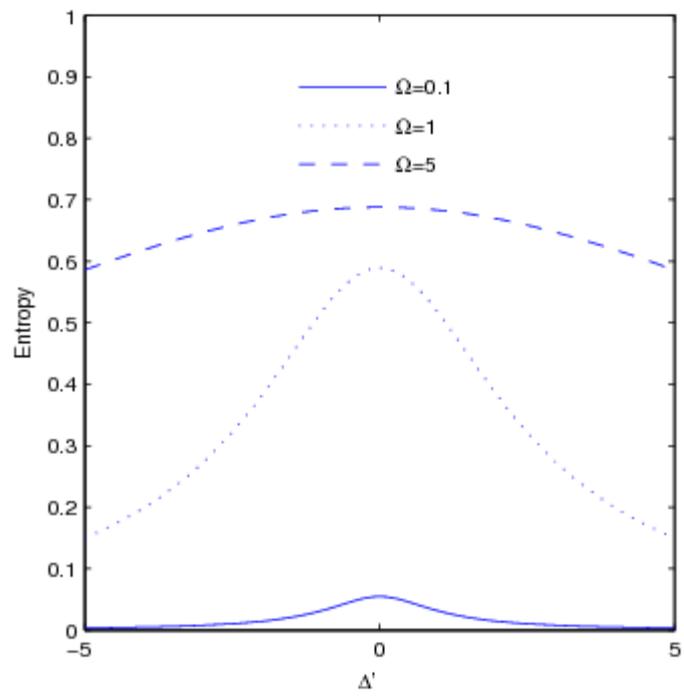

Figure 6